\documentstyle[twoside,epsf]{article}
\catcode`\@=11
\long\def\@makefntext#1{
\protect\noindent \hbox to 3.2pt {\hskip-.9pt  
$^{{\eightrm\@thefnmark}}$\hfil}#1\hfill}               %CAN BE USED 

\def\thefootnote{\fnsymbol{footnote}}
\def\@makefnmark{\hbox to 0pt{$^{\@thefnmark}$\hss}}    %ORIGINAL 
        
\def\ps@myheadings{\let\@mkboth\@gobbletwo
\def\@oddhead{\hbox{}
\rightmark\hfil\eightrm\thepage}   
\def\@oddfoot{}\def\@evenhead{\eightrm\thepage\hfil
\leftmark\hbox{}}\def\@evenfoot{}
\def\sectionmark##1{}\def\subsectionmark##1{}}

\oddsidemargin=\evensidemargin
\renewcommand{\thefootnote}{\fnsymbol{footnote}}
\newcommand{\alphfootnote}
        {\setcounter{footnote}{0}
         \renewcommand{\thefootnote}{\sevenrm\alph{footnote}}}

\newcounter{sectionc}\newcounter{subsectionc}\newcounter{subsubsectionc}
\renewcommand{\section}[1] {\vspace{12pt}\addtocounter{sectionc}{1} 
\setcounter{subsectionc}{0}\setcounter{subsubsectionc}{0}\noindent 
        {\tenbf\thesectionc. #1}\par\vspace{5pt}}
\renewcommand{\subsection}[1] {\vspace{12pt}\addtocounter{subsectionc}{1} 
\setcounter{subsubsectionc}{0}\noindent 
{\bf\thesectionc.\thesubsectionc. {\kern1pt \bfit #1}}\par\vspace{5pt}}
\renewcommand{\subsubsection}[1] {\vspace{12pt}\addtocounter{subsubsectionc}{1}
        \noindent{\tenrm\thesectionc.\thesubsectionc.\thesubsubsectionc.
        {\kern1pt \tenit #1}}\par\vspace{5pt}}
\newcommand{\nonumsection}[1] {\vspace{12pt}\noindent{\tenbf #1}
        \par\vspace{5pt}}

\newcounter{appendixc}
\newcounter{subappendixc}[appendixc]
\newcounter{subsubappendixc}[subappendixc]
\renewcommand{\thesubappendixc}{\Alph{appendixc}.\arabic{subappendixc}}
\renewcommand{\thesubsubappendixc}
        {\Alph{appendixc}.\arabic{subappendixc}.\arabic{subsubappendixc}}

\renewcommand{\appendix}[1] {\vspace{12pt}
        \refstepcounter{appendixc}
        \setcounter{figure}{0}
        \setcounter{table}{0}
        \setcounter{lemma}{0}
        \setcounter{theorem}{0}
        \setcounter{corollary}{0}
        \setcounter{definition}{0}
        \setcounter{equation}{0}
        \renewcommand{\thefigure}{\Alph{appendixc}.\arabic{figure}}
        \renewcommand{\thetable}{\Alph{appendixc}.\arabic{table}}
        \renewcommand{\theappendixc}{\Alph{appendixc}}
        \renewcommand{\thelemma}{\Alph{appendixc}.\arabic{lemma}}
        \renewcommand{\thetheorem}{\Alph{appendixc}.\arabic{theorem}}
        \renewcommand{\thedefinition}{\Alph{appendixc}.\arabic{definition}}
        \renewcommand{\thecorollary}{\Alph{appendixc}.\arabic{corollary}}
        \renewcommand{\theequation}{\Alph{appendixc}.\arabic{equation}}
        \noindent{\tenbf Appendix \theappendixc #1}\par\vspace{5pt}}
\newcommand{\subappendix}[1] {\vspace{12pt}
        \refstepcounter{subappendixc}
        \noindent{\bf Appendix \thesubappendixc. {\kern1pt \bfit #1}}
        \par\vspace{5pt}}
\newcommand{\subsubappendix}[1] {\vspace{12pt}
        \refstepcounter{subsubappendixc}
        \noindent{\rm Appendix \thesubsubappendixc. {\kern1pt \tenit #1}}
        \par\vspace{5pt}}

\newcommand{\textlineskip}{\baselineskip=13pt}
\newcommand{\smalllineskip}{\baselineskip=10pt}

\newcommand{\copyrightheading}[1]
        {\vspace*{-2.5cm}\smalllineskip{\flushleft
        {\footnotesize Quantum Information and Computation, 
          Vol.~2, No.~6 (2002) 434--442 #1}\\
        {\footnotesize \copyright\kern2pt Rinton Press}\\
         }}

\newcommand{\publisher}[2]{{\begin{center}\footnotesize\smalllineskip 
        Received #1\\
        Revised #2
        \end{center}
        }}

\def\abstracts#1#2#3{{
        \centering{\begin{minipage}{4.5in}\footnotesize\baselineskip=10pt
        \parindent=0pt #1\par 
        \parindent=15pt #2\par
        \parindent=15pt #3
        \end{minipage}}\par}} 

\def\keywords#1{{
        \centering{\begin{minipage}{4.5in}\footnotesize\baselineskip=10pt
        {\footnotesize\it Keywords}\/: #1
         \end{minipage}}\par}}
\def\communicate#1{{
        \centering{\begin{minipage}{4.5in}\footnotesize\baselineskip=10pt
        {\footnotesize\it Communicated by}\/: #1
         \end{minipage}}\par}}

\renewenvironment{thebibliography}[1]
        {\frenchspacing
         \ninerm\baselineskip=11pt
         \begin{list}{\arabic{enumi}.}
        {\usecounter{enumi}\setlength{\parsep}{0pt}     
         \setlength{\leftmargin 12.7pt}{\rightmargin 0pt}%FOR 1--9 ITEMS
         \setlength{\itemsep}{0pt} \settowidth
        {\labelwidth}{#1.}\sloppy}}{\end{list}}

\newcounter{itemlistc}
\newcounter{romanlistc}
\newcounter{alphlistc}
\newcounter{arabiclistc}

\newcommand{\fcaption}[1]{
        \refstepcounter{figure}
        \setbox\@tempboxa = \hbox{\footnotesize Fig.~\thefigure. #1}
        \ifdim \wd\@tempboxa > 5in
           {\begin{center}
        \parbox{5in}{\footnotesize\smalllineskip Fig.~\thefigure. #1}
            \end{center}}
        \else
             {\begin{center}
             {\footnotesize Fig.~\thefigure. #1}
              \end{center}}
        \fi}

\newcommand{\tcaption}[1]{
        \refstepcounter{table}
        \setbox\@tempboxa = \hbox{\footnotesize Table~\thetable. #1}
        \ifdim \wd\@tempboxa > 5in
           {\begin{center}
        \parbox{5in}{\footnotesize\smalllineskip Table~\thetable. #1}
            \end{center}}
        \else
             {\begin{center}
             {\footnotesize Table~\thetable. #1}
              \end{center}}
        \fi}

\def\pmb#1{\setbox0=\hbox{#1}
        \kern-.025em\copy0\kern-\wd0
        \kern.05em\copy0\kern-\wd0
        \kern-.025em\raise.0433em\box0}

\def\fnt#1#2{\footnotetext{\kern-.3em
        {$^{\mbox{\scriptsize #1}}$}{#2}}}

\def\fpage#1{\begingroup
\voffset=.3in
\thispagestyle{empty}\begin{table}[b]\centerline{\footnotesize #1}
        \end{table}\endgroup}

\def\runninghead#1#2{\pagestyle{myheadings}
\markboth{{\protect\footnotesize\it{\quad #1}}\hfill}
{\hfill{\protect\footnotesize\it{#2\quad}}}}
\font\tenrm=cmr10
\font\tenit=cmti10 
\font\tenbf=cmbx10
\font\bfit=cmbxti10 at 10pt
\font\ninerm=cmr9

\font\eightrm=cmr8

\font\sevenrm=cmr7

\def\FigName{figure}%
\newbox\captionbox
\long\def\@makecaption#1#2{%
  \ifx\FigName\@captype
    \vskip\abovecaptionskip
    \setbox\tempbox\hbox{{\figurecaptionfont #1\hskip1em #2}}
        \ifdim\wd\tempbox< 28pc
        \centerline{\box\tempbox}
        \else
        {\figurecaptionfont #1\hskip1em #2\par}
\fi\else
        \setbox\tempbox\hbox{{\tablecaptionfont #1\hskip1em #2}}
        \ifdim\wd\tempbox< 28pc 
        \centerline{\box\tempbox}
        \else
        {\tablecaptionfont #1\hskip1em #2\par}%
        \fi   
 \vskip\belowcaptionskip
 \fi}
\InputIfFileExists{psfig.sty}
{\typeout{^^Jpsfig.sty inputed...ok}}{\typeout{^^JWarning: psfig.sty could be be found.^^J}}
\InputIfFileExists{epsfsafe.tex}
{\typeout{^^Jepsfsafe.tex inputed...ok}}
                        {\typeout{^^JWarning: epsfsafe.tex could not be found.^^J}}
\InputIfFileExists{epsfig.sty}
{\typeout{^^Jepsfig.sty inputed...ok}}{\typeout{^^JWarning: epsfig.sty could not be found.^^J}}
\InputIfFileExists{epsf.sty}
{\typeout{^^Jepsf.sty inputed...ok}}{\typeout{^^JWarning: epsf.sty could not be found.^^J}}%
\def\fps@figure{tbp}
\def\ftype@figure{1}
\def\ext@figure{lof}
\def\fnum@figure{Fig.\ \thefigure}
\def\qed{\hbox{${\vcenter{\vbox{                  %HOLLOW SQUARE
   \hrule height 0.4pt\hbox{\vrule width 0.4pt height 6pt
   \kern5pt\vrule width 0.4pt}\hrule height 0.4pt}}}$}}

\renewcommand{\thefootnote}{\fnsymbol{footnote}}  %USE SYMBOLIC FOOTNOTE

\providecommand{\textdegree}{$^{\circ}$}

\begin{document}
\setlength{\textheight}{8.0truein}    %FOR 2ND PAGE ONWARDS

\runninghead{A practical trojan horse for Bell-inequality-based
  quantum cryptography} {Jan-\AA ke Larsson}

\normalsize\textlineskip
\thispagestyle{empty}

\setcounter{page}{434}

\copyrightheading{}

\vspace*{0.88truein}

\fpage{434}
\centerline{\bf
%%%%%%%%%%%%%%%%%%%%%
%Put in titles here
%%%%%%%%%%%%%%%%%%%%%
A PRACTICAL TROJAN HORSE FOR BELL-INEQUALITY-}
\vspace*{0.035truein}
\centerline{\bf BASED QUANTUM CRYPTOGRAPHY}
\vspace*{0.37truein}
\centerline{\footnotesize 
%%%%%%%%%%%%%%%%%%%%%%%%%%%%%%%%%%%%
%put authors' name and address here
%%%%%%%%%%%%%%%%%%%%%%%%%%%%%%%%%%%%
JAN-\AA KE LARSSON\footnote{Present address: Dept.\ of Math.,
  Link\"oping University, SE-581 83 Link\"oping, Sweden.\\
  \hspace*{3pt}E-mail: {\tt jalar@mai.liu.se}}}
\vspace*{0.015truein}
\centerline{\footnotesize\it MaPhySto, Department of Mathematical
  Sciences, University of Aarhus,}
\baselineskip=10pt
\centerline{\footnotesize\it Ny Munkegade, DK-8000 Aarhus C, Denmark}
\vspace*{10pt}
\vspace*{0.225truein}
\publisher{January 10, 2002}{August 20, 2002}

\vspace*{0.21truein} \abstracts{Quantum Cryptography, or more
  accurately, Quantum Key Distribution (QKD) is based on using an
  unconditionally secure ``quantum channel'' to share a secret key
  among two users. A manufacturer of QKD devices could, intentionally
  or not, use a (semi-)classical channel instead of the quantum
  channel, which would remove the supposedly unconditional security.
  One example is the BB84 protocol, where the quantum channel can be
  implemented in polarization of single photons.  Here, use of several
  photons instead of one to encode each bit of the key provides a
  similar but insecure system. For protocols based on violation of a
  Bell inequality (e.g., the Ekert protocol) the situation is somewhat
  different. While the possibility is mentioned by some authors, it is
  generally thought that an implementation of a (semi-)classical
  channel will differ significantly from that of a quantum channel.
  Here, a counterexample will be given using an identical physical
  setup as is used in photon-polarization Ekert QKD. Since the
  physical implementation is identical, a manufacturer may include
  this modification as a Trojan Horse in manufactured systems, to be
  activated at will by an eavesdropper. Thus, the old truth of
  cryptography still holds: you have to trust the manufacturer of your
  cryptographic device. Even when you do violate the Bell
  inequality.}{}{}

\vspace*{10pt} 
\vspace*{3pt} 
\communicate{H-K Lo \& H Zbinden}

\alphfootnote

\vspace*{1pt}\textlineskip      %) USE THIS MEASUREMENT WHEN THERE IS
\section{Introduction}          %) A SECTION HEADING
\vspace*{-0.5pt}
\noindent
%%%%%%%%%%%%%%%%%%%%%%%%%%%%%%%%
%put the text of the paper here
%%%%%%%%%%%%%%%%%%%%%%%%%%%%%%%%
QKD is an application of quantum techniques proposed to give a way of
securely sharing a secret key between two or more users.  The appeal
of this technique is that the security is based on laws of nature
rather than number-theoretical calculations that seem, and perhaps
are, intractable. Progress is quite rapid in this field \cite{GRTZ},
experimental implementations make the necessary technological advances
available \cite{RBGGZ,Daylight} while theoretical analysis provides
better insight into security questions under more and more general
conditions \cite{Mayers1996,ShorPres2000,NPWBK}. And recently (spring
2002), commercial QKD systems have become available. Let us suppose
that Alice and Bob have bought such a commercial QKD system, intending
to use it to share a crypto key secretly from Eve.  Now, it is
possible that they have bought a system which claims to be and behaves
similarly to a QKD system but actually is a (semi-)classical key
distribution system. It would then not provide any security of the
sort inferred from a proper QKD system. Let us look at an example.

\begin{figure}[htbp]
  \centering \psfig{file=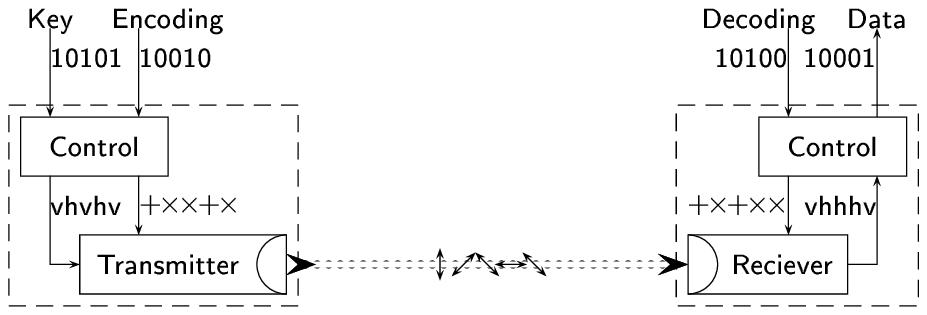} \fcaption{An example of Quantum
    Key Distribution: BB84. Alice generates two random bit-sequences,
    one as the basis of the secret key, and another as the encoding
    for the secret key in the quantum channel.  Alice encodes each bit
    of the key, here in the polarization state of a photon, following
    a certain scheme: If the corresponding encoding-bit is 0, Alice
    uses $0=\mbox{horizontal}$, and $1=\mbox{vertical}$, whereas if
    the encoding-bit is 1, Alice uses the same encoding in a
    45\textdegree\ rotated frame.  The photons are transmitted to Bob
    who uses a third random bit-sequence to decode the bit-sequence
    with the same scheme as Alice.  This means that Bob will have used
    the same de/encoding as Alice only half of the time, but by
    communicating the \emph{settings} used to each other via an
    unjammable classical channel, they can establish for which bits
    they used the same settings. These bits can then be used as
    cryptographic key.}
\end{figure}

In the BB84 protocol \cite{BB84}, security is based on transmitting
the key from Alice to Bob through a quantum channel in such a way that
attempts to eavesdrop is directly detectable as an increased
noise-level in the transmitted key (see Figure~1). The security is
here provided by the quantum-mechanical nature of the photon. Eve
cannot use a beamsplitter to tap off part of the signal for herself
(the so-called ``Beamsplitter attack''); a single photon cannot be
split in such a manner that Bob doesn't notice \cite{Clauser74}. Eve
cannot either faithfully copy the polarization state of the photon,
due to the no-cloning theorem \cite{WootZure}. Lastly, because of the
encoding scheme described above, Eve cannot measure the polarization
and then retransmit another photon to Bob (the ``Intercept-resend
attack''). If she tries, there will be an increased error rate in the
data received by Bob. Alice and Bob will detect this by sacrificing a
portion of the key to estimate the error rate. Then, if the key is not
too noisy, Alice and Bob can use classical privacy amplification
\cite{BBR,BBCM}. Otherwise, they will be forced to abandon the QKD as
being compromised.

This provides good security given that the quantum channel really is
quantum. By constructing the device used by Alice to send several
photons for each key bit, the key can (somewhat simplified) be
successfully extracted by Eve using the Beamsplitter attack. This will
not introduce noise in Bob's data, and thus, there is no security
provided by such a system \cite{BBBSS}.  Many devices used in
practical implementations does not use single photons but weak laser
pulses for which less than one photon is present in each pulse on the
average, and present research indicates that these devices provide the
expected security only under certain restrictions \cite{BLMS}.

\begin{figure}[htbp]
  \centering
  \psfig{file=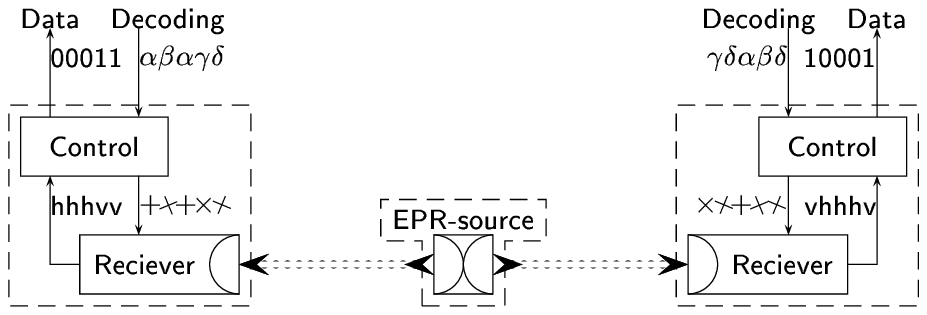}  
  \fcaption{The Ekert protocol: The quantum
    channel contains a bidirectional source and a random key is
    generated at detection at Alice and Bob. In the
    photon-polarization case, the protocol is as follows: the source
    sends one photon to Alice and one to Bob in an entangled state, an
    EPR-Bohm pair \cite{EPR,Bohm51}, which is such that equally
    oriented polarization measurements yield identical results.  Both
    receivers have, in this case, four settings
    ($\alpha=0$\textdegree, $\beta=22.5$\textdegree,
    $\gamma=45$\textdegree, $\delta=67.5$\textdegree) and Alice an Bob
    use random independent settings at their respective sites.  Again,
    by communicating on an unjammable classical channel, they can
    establish for which bits the settings were the same. These bits
    provide the key.}
\end{figure}

Here, we will concentrate on another system, less discussed in this
context: the Ekert protocol \cite{Ekert91} (see Figure~2). Security is
here provided by the quantum-mechanical properties of entangled photon
pairs. In this protocol there is less probability that the settings
are the same, but there is a compensation since one does not need to
sacrifice key bits for eavesdropper detection. One uses the left-over
bits instead, whose correlation provides a violation of a Bell
inequality \cite{Bell64}. The correlation is
\begin{equation}
  E(\phi,\varphi)=\frac%
  {N_{\mbox{same}}(\phi,\varphi)-N_{\mbox{different}}(\phi,\varphi)}%
  {N_{\mbox{same}}(\phi,\varphi)+N_{\mbox{different}}(\phi,\varphi)},
\end{equation}
where $\phi$ and $\varphi$ are the settings at Alice and Bob,
respectively, and $N_{\mbox{same}}$ and $N_{\mbox{different}}$ are the
number of photon pairs for which the results are the same and
different, respectively. It is known that an EPR-Bohm pair
\cite{EPR,Bohm51} yields correlations that violate the CHSH form of
the Bell inequality \cite{CHSH}:
\begin{equation}
 |E(\alpha,\beta)+E(\gamma,\beta)|
 +|E(\gamma,\delta)-E(\alpha,\delta)|\leq2.\label{eq:CHSH} 
\end{equation}
If the difference between Alice's and Bob's setting is
22.5\textdegree\ (as in $E(\alpha,\beta)$, $E(\gamma,\beta)$ and
$E(\gamma,\delta)$), the correlation between the results is
$1/\sqrt2\approx0.7071$, and if the difference is 67.5\textdegree\ (as
in $E(\alpha,\delta)$), the correlation is $-1/\sqrt2$. This yields a
left-hand side of $2\sqrt2$ in (\ref{eq:CHSH}), i.e., a violation of
the CHSH inequality. An attempt to eavesdrop will establish
EPR-elements of reality, and by the local random choice of settings at
the detection sites, the resulting correlation must obey the
inequality (\ref{eq:CHSH}). Thus, the eavesdropper's presence can be
detected by checking the violation \cite{Ekert91}. If it is
sufficient, classical privacy amplification can be used, but in the
worst case, the QKD must be abandoned as being compromised.

\section{A Trojan Horse}

It would seem that violation of the Bell inequality ensures that a
quantum channel really is used, since by the violation, there
\emph{can be no pre-existing key that can be eavesdropped upon}.
Unfortunately, this is not entirely true. There are situations which
allow for a pre-existing key and random local choices of the settings
\emph{while still violating the Bell inequality} (formally). This is
related to the so-called detector-efficiency loophole discussed by
many authors (e.g.\ \cite{GRTZ,Rowe01,Jalar98a,Jalar99a} and
references therein). In what follows, a ``cheat'' will be outlined,
using the same physical setup as in the true Ekert QKD sketched above,
only changing the ``Control'' boxes of Figure~2. The reason for using
the word ``cheat'' is that it is unlikely that a manufacturer would
make a device according to the below specification while believing
that it is a true QKD system (but see the tempting ``Really cheap KD''
in Appendix A).

The first task will be to choose the (semi-)classical channel to be
used instead of the polarization state of each photon. Let us use the
timing of events as our ``hidden variable''\footnote{The recent claim
  made by Hess and Phillip \cite{HessPhila,HessPhilb,HessPhilc}, that
  time dependence has been neglected in the Bell theorem(s) may seem
  similar to this ``cheat''. However, their claim is stronger and more
  controversial, since they claim their model works with ideal
  detectors. Their construction is quite different than the one used
  here, and by all evidence their model is non-local.  In other words,
  it cannot be used for the present task; it cannot violate the Bell
  inequality while being separated into the three ``boxes'': Source,
  Alice's detector, and Bob's detector, unless the settings are
  communicated {\em before} the measurements.}\ \ (this will be
especially simple if we use a pulsed variant of the common parametric
down-conversion source \cite{KMWZSS}). At the start of the protocol,
the devices are initiated so that they are in sync, say by a bright
light-pulse from the source. After this initiation, let us establish a
numbering of time slots, say 8000 short time slots (repeated if
necessary). The controlling electronics is now changed so that instead
of using the polarization measurement result, the received time slot
is mapped into the translation tables in Figure~3 to obtain a
measurement result.

\begin{figure}[htbp]
  \centering
  a)\quad\psfig{file=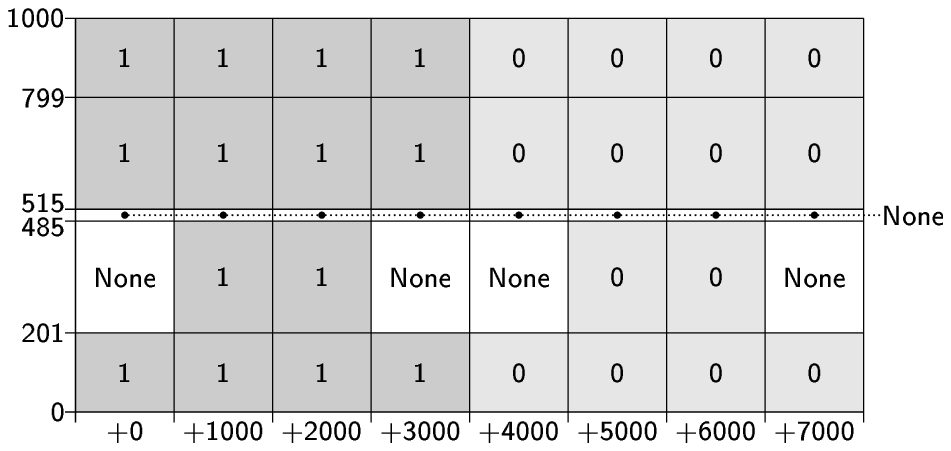}\\
  b)\quad\psfig{file=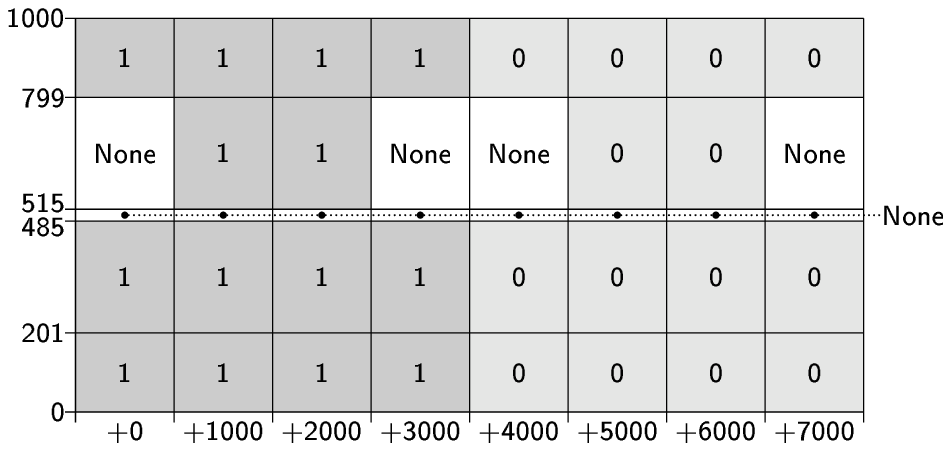}\\
  \fcaption{Translation tables from time slot number to measurement
    result to be used for the $\alpha=0$\textdegree setting. Each
    22.5\textdegree\ shift corresponds to a shift of 1000 in the
    numbering of the time slots (wrapped if it exceeds 8000).  a)
    Translation at Alice's detector.  For example, a photon arriving
    at Alice in time slot 2223 would give the result ``1'' at the
    setting $\alpha=0$\textdegree. Had the setting been
    $\beta=22.5$\textdegree, the controlling electronics would have
    shifted the slot to $2223+1000=3223$, and there would have been no
    detection reported. Similarly, there would have been no detection
    at $\gamma=45$\textdegree, and a ``0'' at
    $\delta=67.5$\textdegree.  b) Translation at Bob's detector.}
\end{figure}

Some received photons will register as ``None'' in the table, and will
not be reported as bits in the output data, so the resulting
efficiency of the detector array will be lowered somewhat. If the
photon pairs are evenly distributed over the 8000 (repeated) slots,
the overall rate of reported detections will be decreased somewhat. Of
the 8000 slots, there are $4*201+12*485$ for which there is a result
reported, so
\begin{equation}
  P\big(\mbox{ bit reported }\big|\mbox{ single detection }\big)
  =\frac{4*201+12*485}{8000}=0.828.
  \label{eq:eta}
\end{equation}

\begin{figure}[htbp]
  \centering \quad\psfig{file=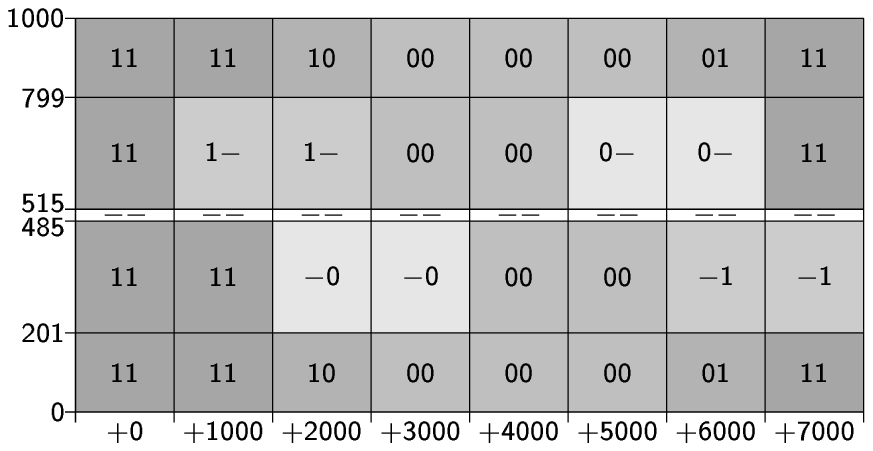} \fcaption{An example of a
    setup where Alice has chosen the $\beta=22.5$\textdegree setting
    and Bob has chosen the $\gamma=45$\textdegree setting. For
    example, a photon arriving at the slot 2810 will give a $1$ at
    Alice and a $0$ at Bob, while a pair arriving in the slot 5560
    will give a $0$ at Alice and no result at Bob.}
\end{figure}

In Figure~4, we can see an example of simultaneous results obtained by
Alice and Bob for a certain setting at each side. Clearly, for a pair
to be reported, the photon pair needs to be detected in a slot where
results are reported on both sides. In a similar fashion as above,
there are $8*201+8*485$ such slots, and thus,
\begin{equation}
  P\big(\mbox{ pair reported }\big|\mbox{ pair detection })
  =\frac{8*201+8*485}{8000}=0.686.
\end{equation}
An important property of these probabilities of single photon
detection and pair detection is that they do not vary with the
settings. Also, these events are approximately statistically
independent since $0.828^2\approx0.686$; the approximation improves
with a larger table. In a realistic situation, if the efficiency is
5\% \cite{WJSWZ} the modified efficiency would be roughly 4\%, which
would not cause great concern in a commercial system. The reader may
recognize (\ref{eq:eta}) as being just below the efficiency bound for
the CHSH inequality.

The obtained results are always the same if the settings are equal. A
knowledgeable user would be concerned with the absence of noise in
this particular implementation, but it is easy to introduce an
appropriate amount of artificial noise in the translation tables of
Figure~3, or use the modified scheme discussed below. If the
``orientations'' differ by $\pm$22.5\textdegree\ (e.g., as in
$E(\alpha,\beta)$), the correlation is
\begin{equation}
  E(\alpha,\beta)=\frac{(4*201+8*485)-4*201}{8*201+8*485}
  =\frac{485}{686}\approx 0.7070.
\end{equation}
The ``error'' relative to the true QM value $1/\sqrt2$ is
approximately $1.1*10^{-4}$, which can be ``improved'' by using a
larger table. A similar calculation yields
$E(\alpha,\delta)\approx-0.7070$, and these correlations will yield a
violation of the CHSH inequality. Even when there are EPR-elements of
reality (there are here!) and the settings are chosen locally and at
random.

While this is not a quantum system, the obtained statistics mimics
one: the Ekert protocol will give the users \emph{an insecure key}
while \emph{the Bell inequality will be violated}.  An eavesdropper
needs not, in this case, access the quantum channel since simple
knowledge of the translation tables together with the public
communication between Alice and Bob will give Eve all she needs to
obtain the key.

One obvious shortcoming of this simple setup is that the results will
show a strong correlation to the time slot number in which the result
was obtained, and detection or no-detection will show a weak
correlation. Also, the devices do not need to be oriented manually
when the system is first set up. In real QKD, at least in this
implementation, the bits correspond to some planar polarization state
of a photon. The cheat may now be slightly modified to use
polarization for one of the four settings, letting the result keep or
invert the translation table, depending on if the result was
``horizontal'' or ``vertical''. This would require manual orientation
of the detectors, and would altogether remove the result correlation
to the time slot number. In addition, it would also require Eve to do
a simplified intercept-resend on the quantum channel.

\section{Discussion}

It is possible to continue the discussion of these and other tests and
ways to counter them, but our space is limited. More importantly, such
tests may not reveal any misbehavior from the receiving devices. This
is because the physical implementation described above is identical to
the one used in true QKD, only changing the mapping from measurement
results to output data.  The devices may operate in ``QKD mode'' when
purchased and provide a true QKD system, but e.g.\ when fed a certain
initiation pulse, change into ``Trojan mode''. The users does not
notice this, since the data they obtain follow the same statistics
(even the lowered efficiency of the Trojan mode can be masked by a
permanent artificial lowering of the efficiency in QKD mode). An
eavesdropper may now ``tap'' the key distribution scheme by inserting
appropriate devices that emit this special initiation pulse when the
normal initiation pulse is received from the source. Having switched
the receptors of Alice and Bob into Trojan mode, the eavesdropper can
listen to their conversation unhindered. And devices like this can be
changed from QKD mode to Trojan mode at will by the eavesdropper.

In principle, a Trojan of this type can be identified but that would
require reverse-engineering the receiving devices.  A complete
dissemination of the internals, including the content of any ROM or
flash RAM present, will show whether the system is a true QKD
implementation or if it does contain a Trojan of the above type. It is
not enough to examine the device superficially, since the Trojan is
such that only the internals (the program) of the ``Control'' is
changed (see Figure~2). This will unfortunately be difficult for a
normal user. Furthermore, one-chip devices, normally intended to make
the devices cheaper, will make it difficult even for a specialist. Of
course, a successful extraction of the key while still violating the
Bell inequality would be direct proof of a Trojan, but one would need
to activate the Trojan first, and this is not easy since the
activation (if it exists) is unknown to Alice and Bob.

Thus, to determine whether a system is true QKD or not will require
advanced testing, and users of commercial QKD devices cannot generally
be expected to have access to the technology needed. Indeed, if they
did, they would be able to build their own QKD system rather than
buying one. It would be possible to defer the tests to a certifying
authority, but in the present world, perhaps this is not a great
improvement: a potential user would then need to trust the certifying
authority rather than the manufacturer.

It would naturally be desirable to close the detector-efficiency
loophole in QKD implementations. The recent result by Rowe
\emph{et~al} \cite{Rowe01} shows promise, but is not yet usable in QKD
setups.  Furthermore, some present implementations use Franson
interferometry \cite{GRTZ,Franson91} in which the quantum channel is
established using time-energy entanglement instead of the polarization
entanglement used here.  And there is a local realistic model of the
Franson setup even in the ideal case \cite{Jalar99c}. In other words,
the model works even at 100\% efficiency, and it would be simple to
use this this model to create a Trojan for the corresponding QKD
scheme, along the lines indicated here.

In the case of BB84 QKD, it has been suggested that a self-checking
source can be constructed via a certain experimental setup and a Bell
inequality test \cite{MayerYao}. The considerations here should be
taken into account when testing and using such a source. For instance,
one may consider buying the source and the testing equipment from
different manufacturers (but this would be only slightly better than
buying both from one manufacturer).

True QKD does have many good properties, and the Trojan described does
not remove any of these good properties. However, users must make sure
that the system they intend to use really is a true QKD system.
Ultimately, the old truth of cryptography still holds: you need to
trust the manufacturer of your cryptographic device. Even when you do
violate the Bell inequality.

\nonumsection{Acknowledgments}

This work was supported by the Wenner-Gren foundation and the Royal
Swedish Academy of Science.  MaPhySto, The Centre for Mathematical
Physics and Stochastics, is funded by the Danish National Research
Foundation.

\nonumsection{References}

\appendix{: Really Cheap KD}

\noindent
A really cheap key-distribution system that pretends to be an Ekert
QKD system can be built out of the following components:
\begin{description}
\item[Source:] Two pulse transmitters of the type commonly used in
  fiber-optic communications, controlled to transmit pulses
  simultaneously at random moments in time. In addition, some
  provision should be made for the initial time sync.
\item[Receivers:] Suitable receiving devices, again of the type
  commonly used in fiber-optic communications, controlled by
  electronics that use the translation tables of Figure~3 after having
  established the time sync.
\end{description}
This system will be cheap, fast, and efficient, compared to a
single-photon system. This is because it can be built out of existing
standard components, the speed is only limited by the specifications
of the components, and the output efficiency is that of the
translation tables. The system will yield a key and violate the Bell
inequality. For a manufacturer, the above construction is tempting.
Violation is all that is needed for secure QKD, is it not? \emph{No.
  This is not a QKD system.} Do not build/buy a system like this. It
is cheap, but you get what you pay for: \emph{no security}.

\end{document}